\documentclass[a4paper]{article}

\usepackage{amsmath}
\usepackage{graphicx}
\usepackage{authblk}
\usepackage{breakcites}
\usepackage{natbib}
\usepackage{tabularx}
\usepackage[font=small]{caption}
\newcommand{\mockalph}[1]{}

\title{Statistical interpretations and new findings on Variation in Cancer Risk Among Tissues}

\date{}

\author[1]{Robert Noble}
\author[1]{Oliver Kaltz}
\affil[1]{Institut des Sciences de l'Evolution, Universit\'{e} Montpellier II, Place E Bataillon, 34095 Montpellier Cedex 5, France}
\author[1,2]{Michael E Hochberg\thanks{Corresponding author. Email: michael.hochberg@univ-montp2.fr}}
\affil[2]{Santa Fe Institute, 1399 Hyde Park Road, Santa Fe, NM 87501, USA}

\begin{document}
\maketitle

\begin{abstract}
\citet{Tomasetti2015} find that the incidence of a set of cancer types is correlated with the total number of normal stem cell divisions. Here, we separate the effects of standing stem cell number (i.e., organ or tissue size) and per stem cell lifetime replication rate. We show that each has a statistically significant and independent effect on explaining variation in cancer incidence over the 31 cases considered by Tomasetti and Vogelstein. When considering the total number of stem cell divisions and when removing cases associated with disease or carcinogens, we find that cancer incidence attains a plateau of approximately 0.6\% incidence for the cases considered by these authors. We further demonstrate that grouping by anatomical site explains most of the remaining variation in risk between cancer types. This new analysis suggests that cancer risk depends not only on the number of stem cell divisions but varies enormously ($\sim$10,000 times) depending on the stem cell's environment. Future research should investigate how tissue characteristics (anatomical site, type, size, stem cell divisions) explain cancer incidence over a wider range of cancers, to what extent different tissues express specific protective mechanisms, and whether any differential protection can be attributed to natural selection.
\end{abstract}

\section*{Introduction}

\citet{Tomasetti2015} (hereafter T\&V) compiled data on 31 cancer types to assess the variation in incidence explained by random factors, as opposed to environmental and inherited factors. They quantified random factors as the total lifetime number of stem cell divisions per tissue or organ type, assuming that cancer has a fixed probability of emerging per stem cell division, and therefore that tissues with more total cell divisions should be more prone to cancer due to stochastic mutations, or ``bad luck". This analysis uncovered a strong positive association between cancer incidence and the lifetime number of stem cell divisions, indicating that random mutations due to replication and repair errors could explain a large part of the variance in risk among cancer types (Fig. 1 in \citet{Tomasetti2015}). Furthermore, the authors quantified the contribution from external environment and inherited factors by an Extra Risk Score (ERS). Cancers with high ERS are indeed known to be associated with carcinogenic exposure (Fig. 2 in \citet{Tomasetti2015}).

In this article, we reanalyse the dataset of Tomasetti and Vogelstein to reveal more about sources of variation in risk between cancer types. We show that \textit{c.} 50\% of this variation is due to tissue size, indicating that independent of stem cell divisions, larger tissues are more likely to harbor cancers than smaller tissues. A simpler measure of Extra Cancer Risk (a classification of cancers most likely to be caused by carcinogens) yields similar findings to T\&V, with some notable differences. Moreover, when using the total number of stem cell divisions as a metric and only considering cancers that are not typically the result of disease or carcinogenic exposure, we find that cancer incidence plateaus at approximately 0.6\% for the cases in T\&V's dataset. We further demonstrate that most of the remaining variation in cancer risk can be explained by grouping cancers by anatomical site (e.g., pancreas, bone, intestine), and that each site has a very different risk per stem cell division. Our study indicates new directions for research in showing how tissue characteristics may independently explain variation in cancer incidence. We suggest that evolution by natural selection may contribute to the basic pattern.

\section*{Results}

\subsection*{Independent contributions of division rate and stem cell number}

Tomasetti and Vogelstein calculated the total lifetime number of stem cell replications (lscd) as the product of the size of the organ's stem cell population ($s$) and the lifetime number of divisions per stem cell ($d$). They then tested for a correlation between lscd and cancer incidence. One way in which their analysis could be extended is to differentiate the individual contributions of $s$ and $d$ to cancer risk. For instance, a small stem cell population with many replications (e.g., esophageal cells) may have the same lscd as a large stem cell population with few replications (e.g., lung cells, see Table S1 in \citet{Tomasetti2015}). However, in the former case, cancer risk may result mainly from replication error, while the latter has a considerably larger number of cells potentially exposed to carcinogenic environments at any point in time. 

\begin{figure}
  \centering
    \includegraphics[width=\textwidth]{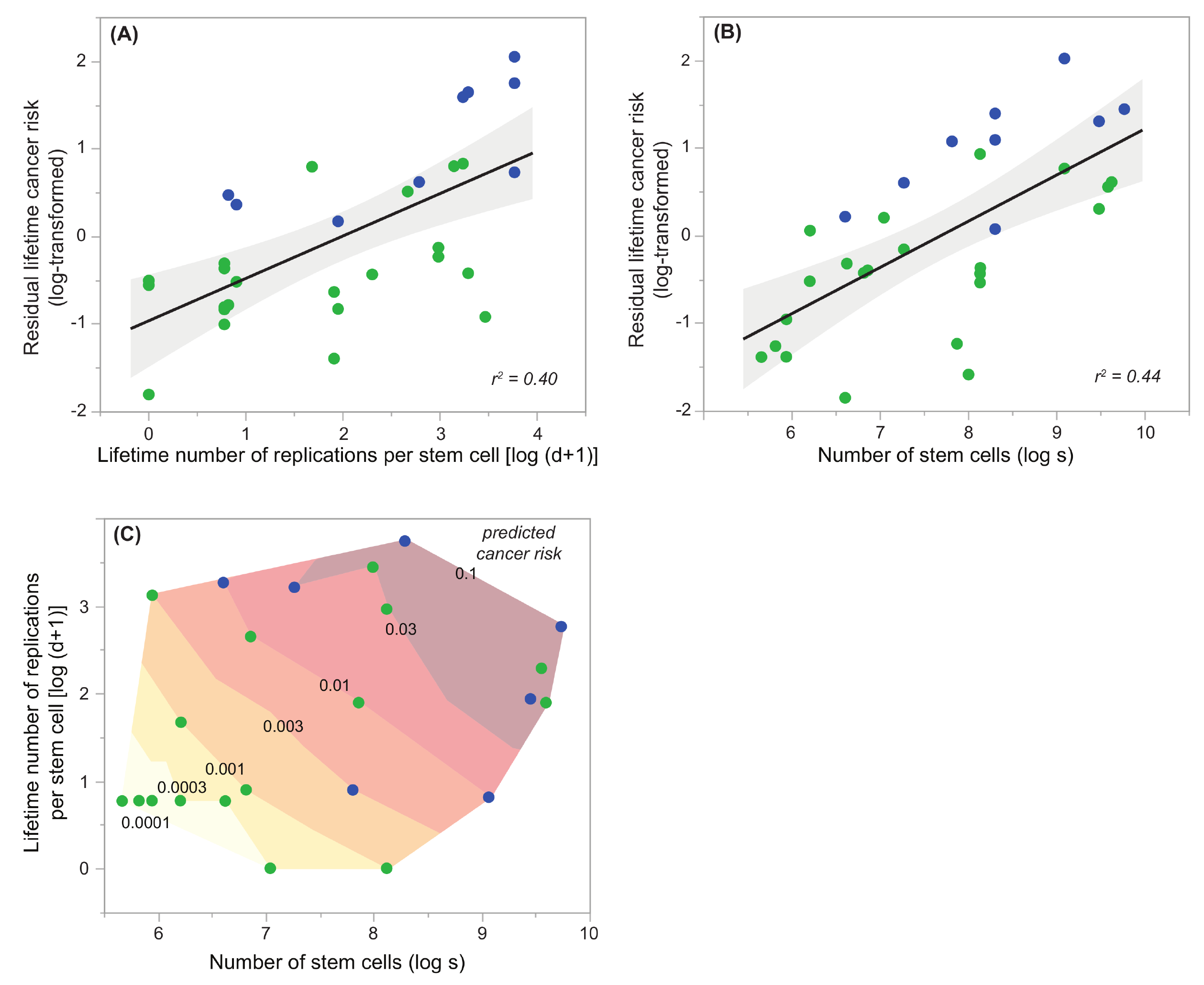}
  \caption{Relationships between the number of stem cells per tissue ($s$), the lifetime number of replications per stem cell in that tissue ($d$) and lifetime cancer risk, across 31 cancer types (data from Table S1 in \citet{Tomasetti2015}). \textbf{A} Relationship between stem cell replication ($d$) and cancer risk, after statistically correcting for the effect of stem cell number ($s$). This correction was done by regressing cancer risk on $s$, and then performing the partial regression of residual cancer risk on $d$. The $r^2$ value is the square of the partial regression coefficient and quantifies the amount of variation in residual cancer risk explained by stem cell division. \textbf{B} Relationship between stem cell number ($s$) and cancer risk, after statistically correcting for the effect of stem cell replication ($d$). The partial $r^2$ quantifies the variation in residual cancer risk explained by stem cell replication. \textbf{C} Illustration of the combined positive effects of stem cell number ($s$) and stem cell replication ($d$) on predicted cancer risk. Predicted values were obtained from the multiple regression of cancer risk on $d$ and $s$ (see text). In 0.5 log-intervals we assigned a colour gradient to the predicted values, ranging from light orange (low predicted risk) to dark red (high predicted risk). Thus, cancer risk increases with increasing values of both $s$ and $d$. All analyses and figures use log-transformation of $s$, $d$ and cancer risk. The black lines in \textbf{A} and \textbf{B} represent regression lines, and the shaded areas the 95\% confidence intervals around the regression. Colour-coding based on Fig. 2 in \citet{Tomasetti2015}, denoting deterministic D-tumours (blue) and replicative R-tumours (green).}
\label{fig:split}
\end{figure}

We conducted a multiple regression analysis with cancer incidence as the response variable, and $\log(d+1)$ and $\log s$ the explanatory variables:
\begin{equation}
\log \text{cancer risk} \sim \log (d+1) \times \log s.
\label{eqn:multi}
\end{equation}
There was no significant correlation between the two explanatory variables $(r = 0.16, n = 31, p > 0.3)$, indicating that variation in $s$ is largely independent of variation in $d$. The multiple regression revealed significant positive effects of both $\log s$ and $\log(d+1)$ on cancer risk $(F_{1, 27} > 18, p < 0.0002)$; the interaction between and $\log s$ and $\log(d+1)$ was not significant $(F_{1, 27} = 0.21, p > 0.6)$. Overall, as expected, the model explained 65\% of the variation in cancer risk, which is identical to the estimate for the composite lscd in \citet{Tomasetti2015}. When correcting for effects of $s$, stem cell division explains 40\% of the variation in risk; conversely, tissue stem cell number explains 44\% of the variation after correcting for per stem cell divisions (Figures~\ref{fig:split}A, B). Figure~\ref{fig:split}C depicts the combined positive effects of $\log(d+1)$ and $\log s$: cancer risk increases with both increasing stem cell number and replication in the organ.

\begin{figure}
  \centering
    \includegraphics[width=0.9\textwidth]{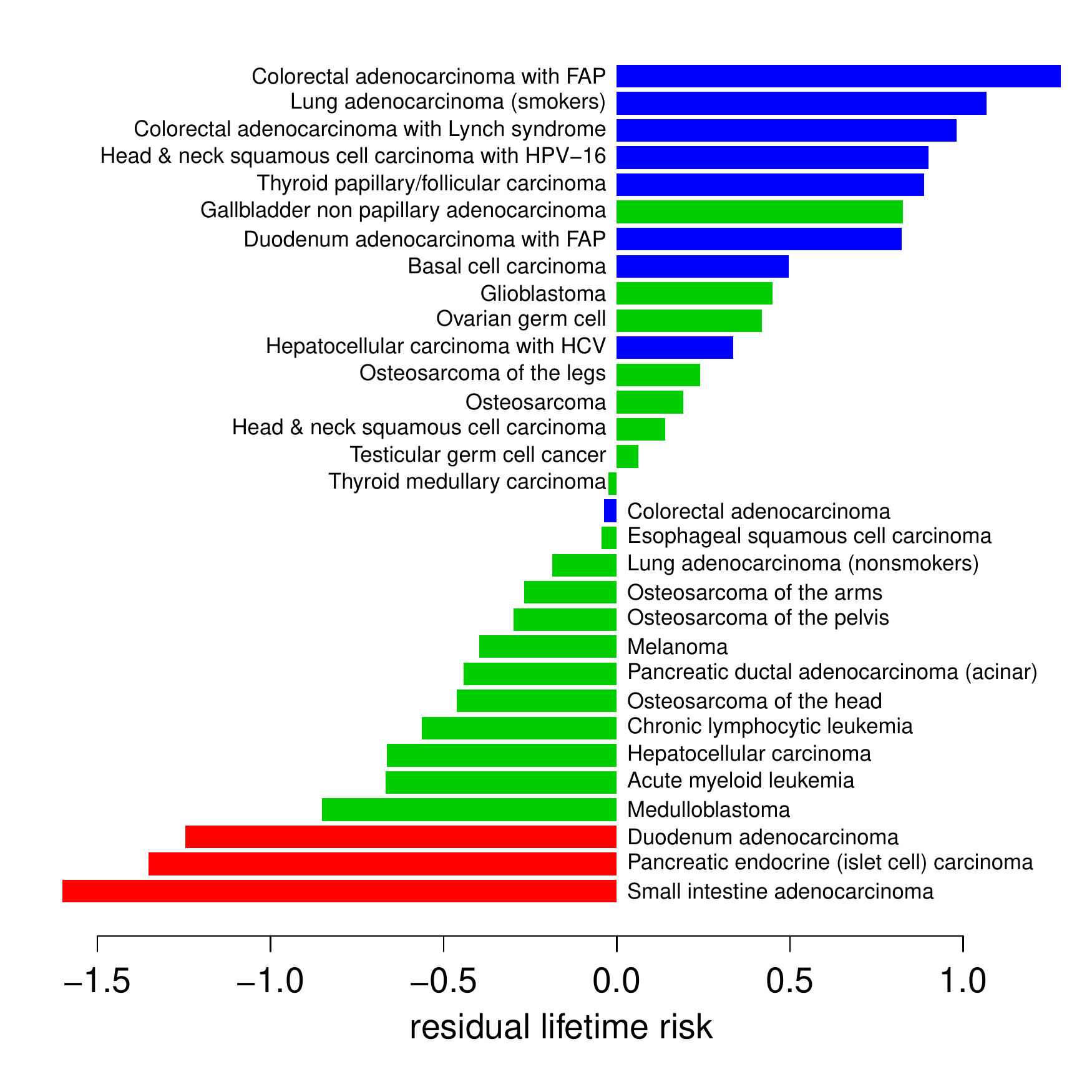}
  \caption{Residual lifetime risk of 31 cancer types, calculated as the difference between observed values and predictions of our multiple regression model (Figures~\ref{fig:split}B, C). Most of the cancers that T\&V classed as deterministic D-tumours (blue bars) also have high residual risks according to our alternative metric. Many such cancers are associated with known causative factors (oncoviruses, chemical carcinogens, or inherited cancer susceptibility genes). The additional identification of cancers with very low residual lifetime risks (red bars) suggests that some tissue types may be differentially resistant to tumours arising from replication events or ``bad luck".}
\label{fig:bars}
\end{figure}

Based on our regression model, we propose a simple alternative evaluation of the replication-independent Extra Cancer Risk (ERS) score. Whereas T\&V calculate the ERS as the product of the logarithms of lifetime risk and total stem cell replications, we use the residual lifetime risk, describing the difference between observed and predicted values from the regression (Figures~\ref{fig:split}B, C). Like the ERS, our more intuitive method identifies a subset of cancers that occur more often than we would expect from the lifetime number of stem cell divisions, including most of those that T\&V classed as deterministic D-tumours (Figure~\ref{fig:bars}, blue bars). Of equal importance for understanding possible causation, the residual lifetime risk also quantifies the extent to which some cancer rates are lower than expected. Carcinomas of the small intestine, duodenum and pancreas are more than ten times less frequent than one would predict from the total number of stem cell divisions (Figure~\ref{fig:bars}, red bars). We note that very similar results can be obtained using the residuals from the regression of risk against lscd, as has been proposed by \citet{Tomasetti2015musings} and \citet{Altenberg2015} since the publication of \citet{Tomasetti2015}.

\subsection*{The saturation of cancer risk}

The relatively shallow gradients of the linear regression models (\textit{c.} 0.5) present a challenge to the hypothesis that the variation in cancer risk is largely due to differences in lifetime numbers of stem cell divisions. If each stem cell division carries the same risk of initiating cancer, then doubling the number of stem cell divisions should correspond to doubling the cancer risk. Therefore the slope of the correlation between cancer risk and lscd should be close to 1. Instead, the gradient of the one-factor linear regression model of T\&V is only 0.53. This discrepency has been noted before \citep{Tomasetti2015musings, Altenberg2015} but has not, in our view, been sufficiently investigated.

\begin{figure}
  \centering
    \includegraphics[width=0.7\textwidth]{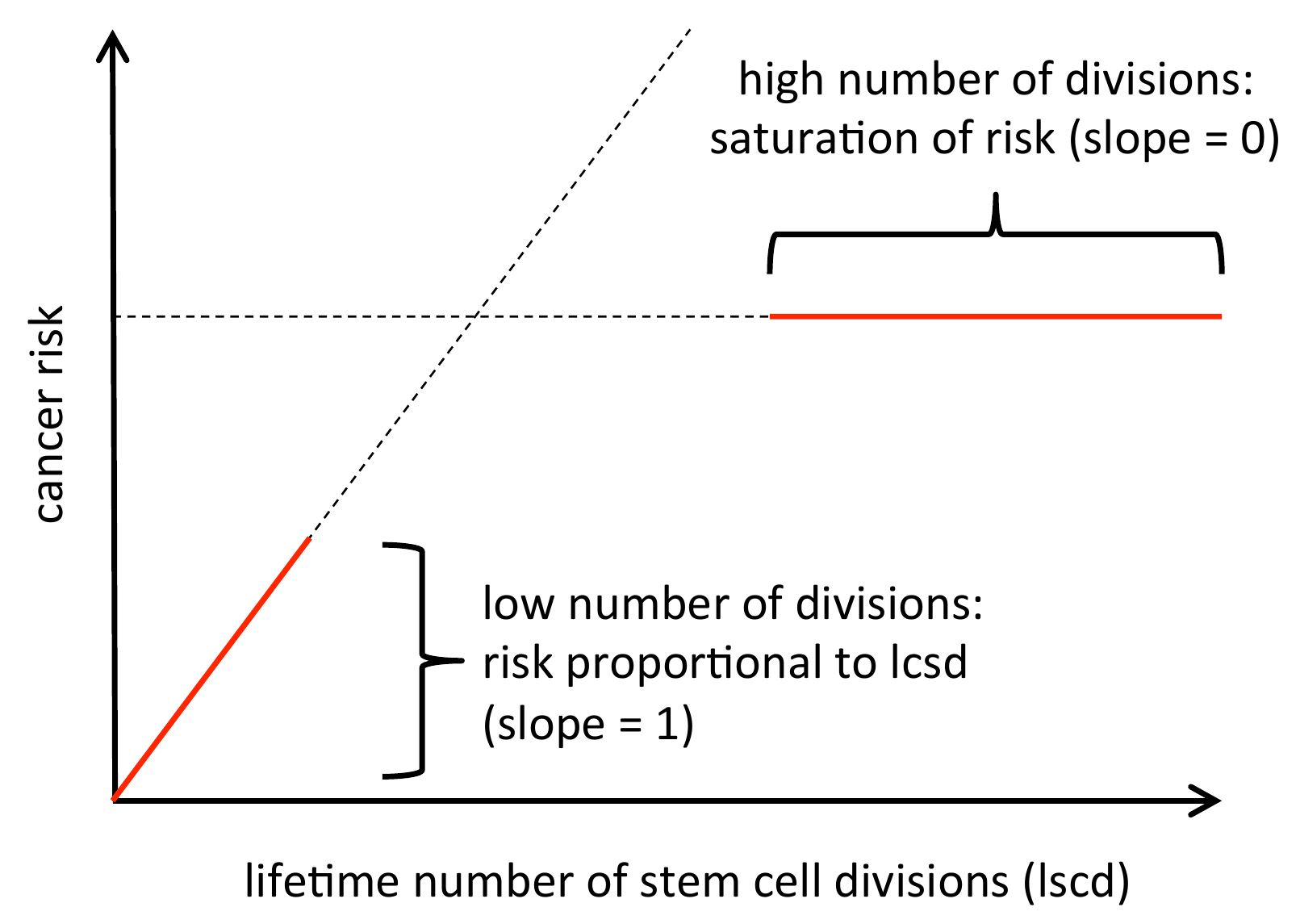}
  \caption{A hypothetical one-factor, non-linear model of the relationship between cancer risk and lifetime number of stem cell divisions (lscd). If each stem cell division has the same probability of causing cancer then there should be a linear relationship between cancer risk and lscd with a gradient of 1. However, we argue that the risk cannot rise indefinitely but must be bounded by a maximum limit, either due to the primacy of other causes of mortality and/or due to differential cancer prevention in tissues with larger total numbers of stem cell divisions.}
\label{fig:explanation1}
\end{figure}

We suggest that the overall gradient is shallower than expected because lifetime risk cannot increase indefinitely, but rather saturates at a maximum level. Mathematically, no risk can exceed 100\%. However, there are two more likely (non-mutually exclusive) reasons for a saturating effect. First, different causes of mortality (e.g., cancers, heart disease, cerebrovascular disease, accidents, etc.) each have a characteristic probability distribution as a function of age. Because of the primacy of mortality events, increases in the probability of a given mortality type will tend to be reflected as \emph{increased} incidence as the age at which the event occurs \emph{decreases}. Thus, all else being equal, a given source of mortality will not exceed approximately $1/ N$, where $N$ is the total number of possible attributed causes of mortality. Of course, all else is not equal, but nevertheless we would expect a saturation effect since the cancers in T\&V's dataset tend to be life threatening at older ages (and therefore have less primacy). Second, to the extent that different tissues are differentially vulnerable to life-threatening cancers, natural selection is expected to result in tissue specific protection \citep{nunney1999lineage}. That all tissues do not employ the same protection mechanisms would be suggestive of either a fitness cost of cancer protection to the organism (i.e. that the cost of added protection in terms of reductions in survival and reproduction outweighs the benefits of lowered risks of life-threatening cancer), or that the phylogenetic emergence of tissue specific protection was somehow linked with tissue differentiation during ontogeny. Therefore, for either or both of the two hypotheses, we would expect the correlation of risk and lscd to have a gradient of 1 only for tissues that have relatively few lifetime stem cell divisions. For high-lscd tissue types we would expect the risk to be bounded by a maximum limit, as illustrated in Figure~\ref{fig:explanation1}.

\begin{figure}
  \centering
    \includegraphics[width=\textwidth]{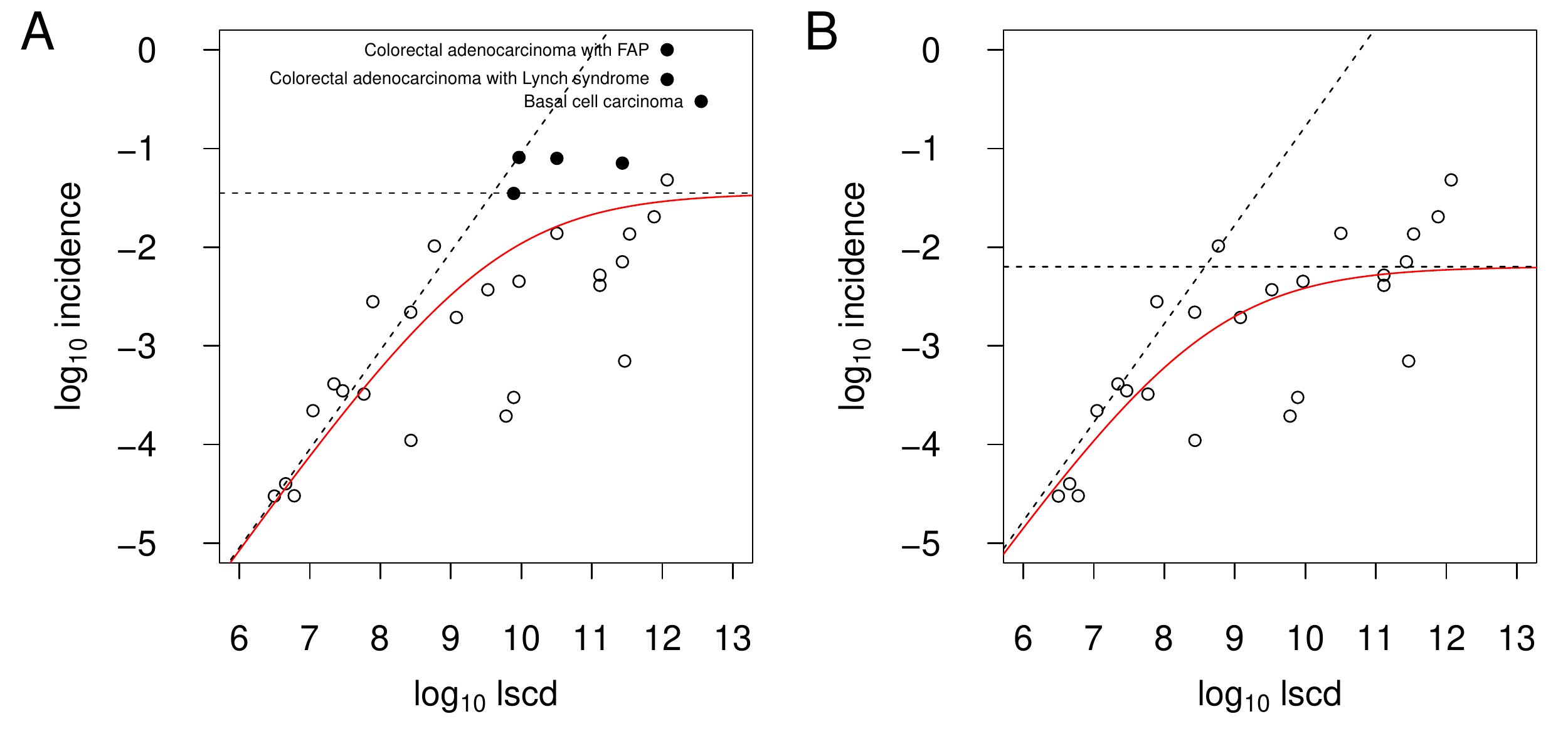}
  \caption{\textbf{A} Relationship between cancer risk and lifetime number of stem cell divisions (lscd) in 31 cancer types, according to a model that assumes that the gradient of the correlation is 1 for small lscd, and is 0 for large lscd. \textbf{B} The same model fitted to the set of 24 cancer types not associated with a high-risk subpopulation (the excluded data points are shown as filled circles in \textbf{A}). The model assymptotes are included as dashed lines.}
\label{fig:logistic}
\end{figure}

A simple model that is consistent with these assumptions is
\begin{equation}
y = -\log(a + e^{-x - b}),
\label{eqn:logistic}
\end{equation}
where $y$ is cancer risk and $x$ is lscd (both log-transformed). For small $x$ this function approaches $y = x + b$ (slope = 1), and for large $x$ it approaches $y = -\log a$ (slope = 0).

Figure~\ref{fig:logistic}A shows the result of fitting the above model to the data for all 31 cancer types. Most of the types with higher than expected risk, according to this model, belong to subpopulations exposed to carcinogens (filled circles in Figure~\ref{fig:logistic}A). These include lung cancer in smokers, intestinal cancer in those with certain inherited genetic alterations, liver and head and neck cancer in those infected with an oncovirus, and basal cell carcinoma, which is generally correlated with a combination of genetic factors and UV-light exposure, and which is very rarely fatal \citep{wong2003basal}.

When the risks related to specific subpopulations are omitted, the lifetime risk per cancer type saturates at around 0.6\% (with s.e.: 0.4-1.1\%, Figure~\ref{fig:logistic}B). Therefore the data appear to be consistent with a model in which the risk of life-threatening cancer increases with lscd with a slope 1, until it is bounded by a threshold of \textit{c.} 0.6\%, i.e., that is well below the theoretical maximum of 100\%. Although the fit of this model is statistically similar to that of the linear model (residual standard errors 0.59 and 0.61, respectively), it is more biologically plausible, and it may therefore reveal more about the multiple factors that determine cancer risk, including natural selection.

\subsection*{Variation between tissues}

We next aim to account for some of the approximately one third of variation that is not explained by our linear and non-linear models. First consider the case of osteosarcoma: the dataset of \citet{Tomasetti2015} includes four types of this bone and joint cancer corresponding to different parts of the body (arms, legs, head, pelvis) and another data point for the entire body. These tissues are thought to derive from similar stem cells in similar environments \citep{bianco2001bone}. Therefore we would expect osteosarcoma risk to increase, with slope 1, according to lscd (or equivalently, with the number of stem cells per tissue, since the division rates are assumed to the same for all osteocarcinomas \citep {spalding2005retrospective}). In fact, the gradient for the five osteosarcomas is 1.25 (s.e. $\pm 0.30$), consistent with the hypothesis.

\begin{figure}
  \centering
    \includegraphics[width=\textwidth]{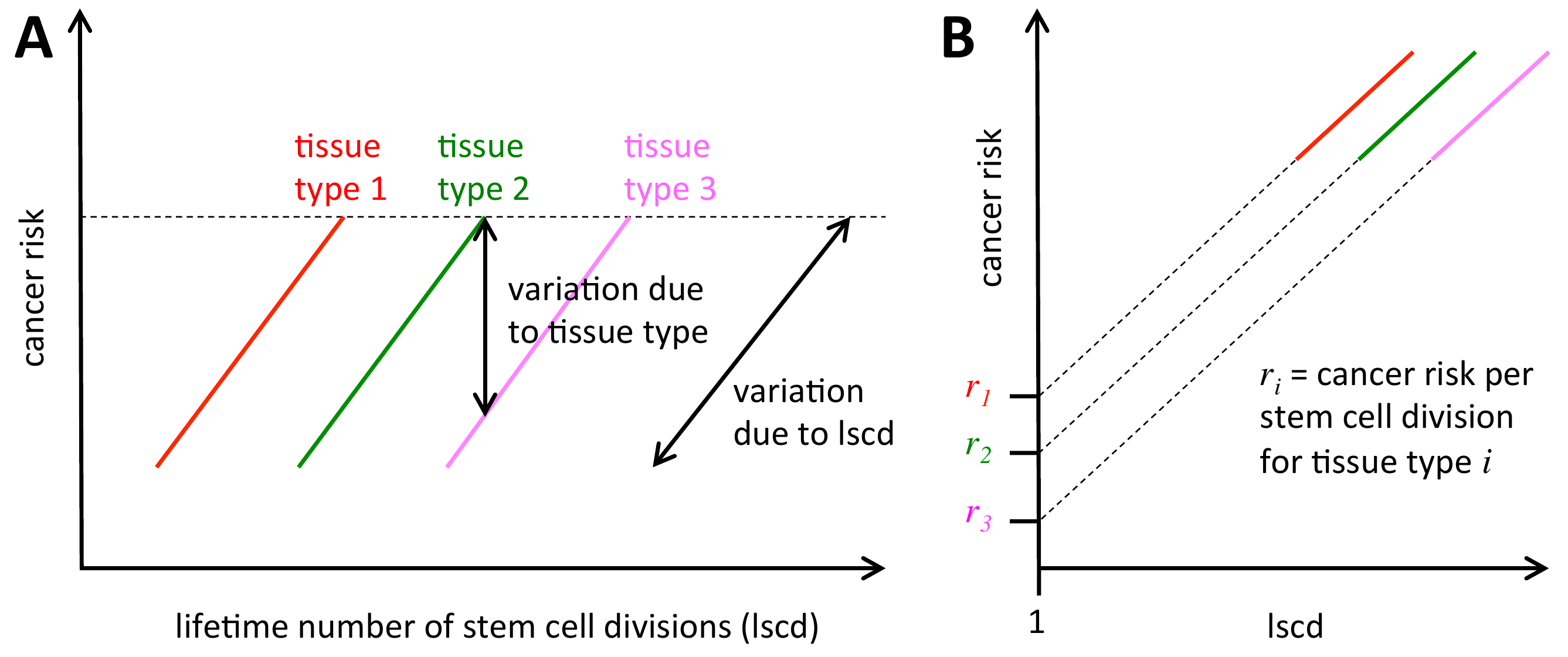}
  \caption{A hypothetical two-factor, linear model of the relationship between cancer risk and lifetime number of stem cell divisions (lscd). In this case the cancer types are divided into subsets according to tissue type. \textbf{A} The subsetting pariations variation into within-subset variation (due to lscd) and between-subset variation (due to tissue type). The gradient of the correlation within subsets is expected to be close to 1. The dashed line indicates a maximum risk threshold. \textbf{B} For each tissue type, the cancer risk per stem cell division can be estimated by extrapolating the regression line to the point where lscd = 1 (i.e. log lscd = 0).}
\label{fig:explanation2}
\end{figure}

Taking this idea one step further, we can split the T\&V data set into subsets of related cancer types and thus divide the variation in risk into two parts (Figure~\ref{fig:explanation2}A). If the members of each subset have the same cancer risk per stem cell divsion then variation within subsets will be mostly due to lscd, whereas variation between subsets will be related to tissue type and/or environment. In this case, we would expect the slope of each subset regression line to be approximately 1. We could then estimate the cancer risk per stem cell division by finding where each regression line intercepts the vertical axis (i.e. log lscd = 0), as illustrated in Figure~\ref{fig:explanation2}B.

\begin{figure}
  \centering
    \includegraphics[width=\textwidth]{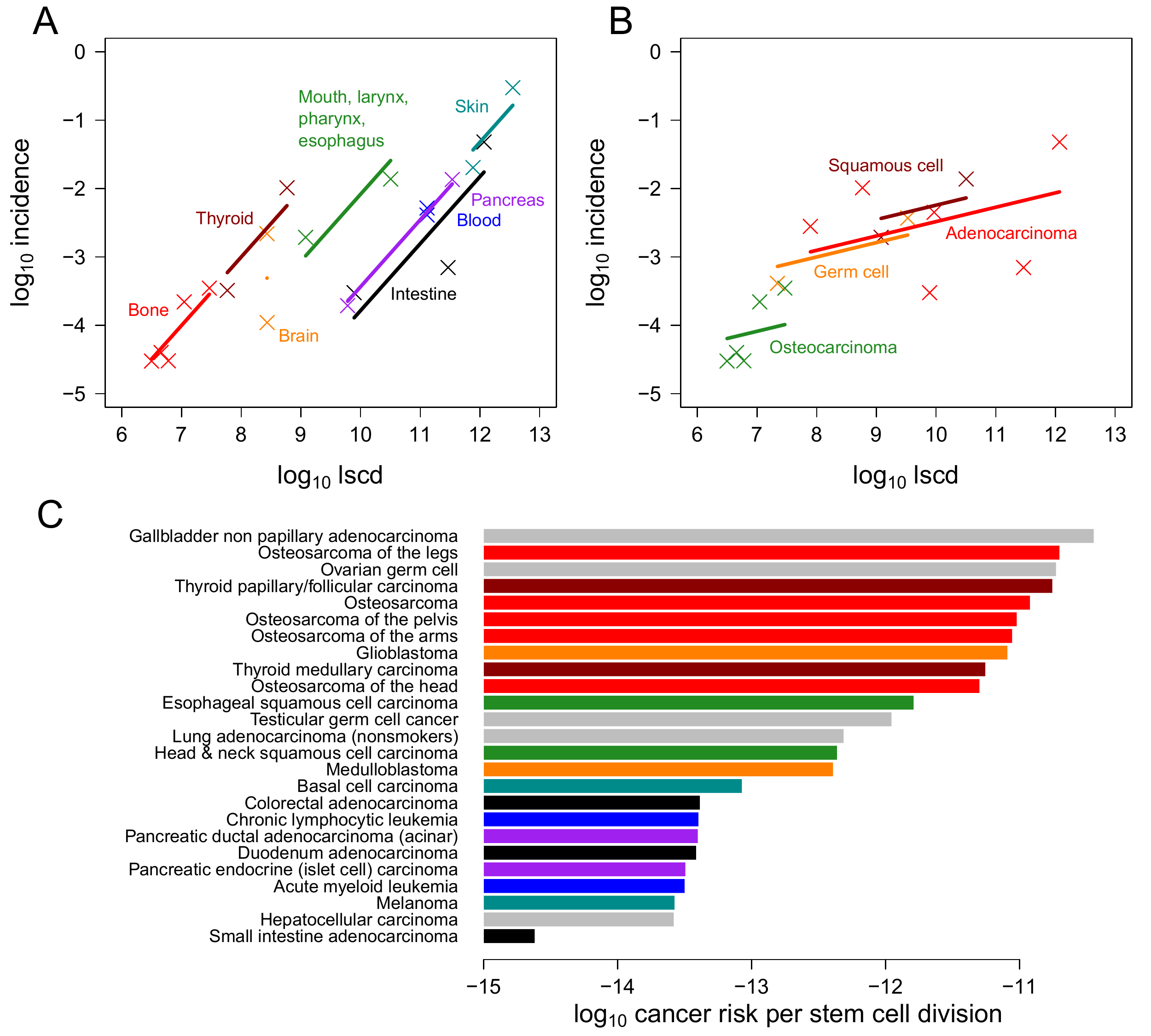}
  \caption{\textbf{A} Relationship between cancer risk and lifetime number of stem cell divisions (lscd) in eight topographically-defined subsets of 20 cancer types (Table~\ref{tbl:subsets_top} in Appendix). The model assumes that the risk per stem cell division may differ between subsets but that the slope of the correlation is the same for each subset. \textbf{B} Relationship between cancer risk and lifetime number of stem cell divisions (lscd) in four morphologically-defined subsets of 15 cancer types (Table~\ref{tbl:subsets_morph} in Appendix). \textbf{C} Cancer risk per stem cell division for 25 cancer types, calculated by dividing risk by lscd. This formula assumes that the correlation between risk and lscd has a gradient of 1 for each tissue type, which is supported by the results of the regression model (Equation~\ref{eqn:logistic}). Cancer types are coloured by topographic subset, according to the scheme shown in \textbf{A}. Five types that belong to topographic subsets with only one member (and so were excluded of the analysis shown in \textbf{A}) are shown in grey.}
\label{fig:sets}
\end{figure}

We chose to define subsets of cancer types according to the widely-used International Classification of Diseases for Oncology (ICD-O) \citep{WHOinternational}. The ICD-O assigns a topographical code for the site of the cancer (e.g. thyroid, skin, lung), corresponding to the environment in which the cancer arises, and a morphological code (e.g. adenocarcinoma, follicular carcinoma, glioblastoma) describing the type of cell. We hypothesised that differences in topography and/or morphology might help explain variation in risk among cancer types.

We first divided the cancer types in the T\&V dataset into topographical subsets, excluding six risk values that apply only to particular subpopulations (Table~\ref{tbl:subsets_top} in Appendix). We then fitted a two-factor regression model to the subsets containing at least two data points (8 subsets, 20 cancer types):
\begin{equation}
\log \text{cancer risk} \sim \log \text{lscd} + \text{subset}.
\label{eqn:subsets}
\end{equation}
This means that we assumed, for each cancer type $i$,
\begin{equation}
\log \text{cancer risk}(i) = A \log \text{lscd}(i) + B(\text{subset}(i)).
\end{equation}
In this model, the slope ($A$) of the linear regression line is assumed to be the same for all subsets, but the intercept ($B$) is allowed to vary depending on the subset. Therefore there are nine parameters (one slope, and eight subset-specific intercepts).

The two-factor regression model explains 90\% of the variation in cancer risk among the 20 cancer types ($F_{8,11} = 12.3$, Figure~\ref{fig:sets}A). Log lscd by itself explains 69\% of the variation, similar to the figure for the full set of 31 analyzed by T\&V, whereas the subset factor explains an additional 21\% (subset effect: $p = 0.04$). In other words, by assuming that the topographical subsets have the same slope but different intercepts, we can explain significantly more of the variation in risk. Moreover, the gradient within the subsets is 0.98 ($\pm 0.19$ s.e.), which is, as predicted, very close to 1.

Note that we chose to include skin cancers in this analysis even though most of the skin cancer risk in the T\&V data set is associated with UV-light exposure \citep{scotto1983incidence}. Since UV-light exposure is assumed to increase the mutation risk per stem cell division, we would expect this environmental factor to shift the regression line for the skin cancer subset upwards, towards higher cancer risk, but we would still expect the slope to be approximately 1. Indeed, the model fit for skin cancer is similar to that for the other subsets (Figure~\ref{fig:sets}A).

Much of the remaining variation is due to the brain cancer subset, but it can be argued that this subset is poorly defined. Whereas glioblastoma is thought to develop in the mature brain, medulloblastoma is believed to originate in the different environment of the early embryo \citep{roussel2011cerebellum}, and it is the only cancer in the T\&V data set that predominantly occurs during childhood (median age 9 years at diagnosis). When the brain cancer subset is excluded, the two-factor regression model explains 93\% of the variation ($F_{7,10} = 19.1$) and the subset factor has a more significant effect ($p = 0.01$).

Apart from brain cancers, there is only one cancer type that substantially deviates from the topographical subsets model: although colorectal and duodenum adenocarcinomas lie almost exactly on a line of slope 1 (also grouping with pancreatic cancers), small intestine adenocarcinoma falls well below this line, being approximately ten times less common than predicted. Therefore a testable prediction of our model is that small intestine adenocarcinoma differs in some important way from the two other intestinal cancers (colorectal and duodenum adenocarcinomas), or that the data for this cancer type is inaccurate.

We also divided the data according to ICD-O morphological code, resulting in four subsets containing at least two data points, which together included 15 cancer types (Table~\ref{tbl:subsets_morph} in Appendix). In the two-factor regression model (Equation~\ref{eqn:subsets}), the morphological subset factor is not significant ($p=0.38$). Therefore we found no evidence that cancer risk in this dataset is related to cell type, independent of anatomical site (Figure~\ref{fig:sets}B). Neverthless, since topography and morphology are moderately correlated in the T\&V dataset, our results do not rule out a combined effect.

Given that the gradient of the correlation between cancer risk and lscd appears to be close to 1 for each topographical type, we can calculate
\begin{equation}
\text{cancer risk per stem cell division} = \text{risk} \div \text{lscd}.
\end{equation}
The estimated risks per stem cell division for each individual cancer type are shown in Figure~\ref{fig:sets}C. These estimated risks vary by nearly four orders of magnitude -- from less than $10^{-14}$ for small intestine adenocarcinoma, to approximately $10^{-11}$ for osteocarcinoma and thyroid cancers -- yet our eight subsets explain 92\% of this variation in twenty cancer types ($F_{7,12} = 18.7, p = 1 \times 10^{-5}$).

Therefore variation in cancer incidence in the dataset of T\&V can be explained by the total number of stem divisions (lscd; \citep{Tomasetti2015}), but can also be understood as variation explained by tissue size and by per stem cell divisions (this study). When using the composite quantity lscd and only considering cancers that are not linked to heredity, disease or mutagenic exposure, we find that anatomical site explains most of the residual variation.

The code used to implement each of our models is provided in an Appendix.

\section*{Discussion}

Despite limitations in the Tomasetti and Vogelstein dataset, it contains a wealth of information that goes beyond their initial analysis. We have made four new findings based on their dataset. First, the total number of stem cells and the lifetime number of divisions per stem cell each significantly, and independently of one another, explain variation in cancer incidence (Figure~\ref{fig:split}). Indeed, our finding of a significant correlation of $s$ with cancer risk is consistent with the prediction that cancer incidence increases with the standing population size of an organ \citep{albanes1988cell, roychoudhuri2006cancer}. One possible mechanism for the tissue size effect is mutations associated with the $2s$ cell divisions during ontogeny for certain tissues \citep{degregori2013challenging}. Whether such mutations are due to random effects is discussed below. Second, our more intuitive measure of Extra Cancer Risk yields results that largely concord with T\&V, but also yield certain notable differences (Figure~\ref{fig:bars}). Third, when assessing a subset of 24 cancers that are not primarily linked to pathogens, disease, or carcinogenic exposure, we find a saturating effect of total stem cell divisions on cancer incidence, with a plateau at about 0.6\% (Figure~\ref{fig:logistic}). This could be explained either by the primacy of mortality events limiting maximal mortality for any single type of event and/or increased cancer prevention mechanisms in tissues with the most total stem cell replications. Fourth, when dividing a subset of 20 cancers by anatomical site, we find that each type shows the same slope of \textit{c.} 1, but is displaced over 4 orders of magnitude in risk per stem cell division, consistent with the hypothesis that different tissues have contrasting protection mechanisms against cancer. We provide testable estimates for each anatomical site of the probability that a single stem cell division will result in cancer. We briefly discuss the implications of these findings below.

Our analysis clarifies one of the main findings of \citet{Tomasetti2015}: there is indeed a direct contribution of stem cell division rate ($d$) to cancer risk, independent of the size of the organ ($s$). This division rate effect is weaker than that found by T\&V, because our analysis is based on replications per stem cell rather than over the population of stem cells. We argue that it is important to distinguish between effects of $d$ and $s$, because stem cells may be replaced by mutated daughter cells following mutagenic exposure in S-phase of the cell cycle \citep{Cairns06082002, branzei2005dna}. Thus, our untested hypothesis is that as exposure to mutagens becomes increasingly chronic, the probability that mutation should correlate with total number of `targets' (that is, with the size of the tissue cell population, and therefore with $s$), whereas, as exposure is increasingly punctual, the probability will correlate with both $d$ and $s$ (i.e., with total stem cell divisions). In either scenario therefore, we suggest that actual `causes' of cancer in Figure 1 of T\&V cannot necessarily be attributed to random mutations only, since these same stem cells may either already carry inherited cancer susceptibility genes, or be replaced by mutated daughter cells exposed to stressful environments. This means that ERS is a conservative estimate of hereditary and environmental causes and that other baseline effects may be indistinguishable from random mutations in the principal dataset, and therefore amplifies T\&V's conclusion of the importance of primary and secondary prevention for a range of cancers \citep{hochberg2013preventive, vogelstein2013cancer}.

When considering the total number of stem cell divisions as a single metric that explains most of the variation in incidence, our study provides evidence for an effect of anatomical site, corresponding to the environment in which cancer arises. The variation between tissues is analogous to Peto's paradox, whereby biological species of larger body mass and/or longer life span exhibit smaller than expected incidences of cancer \citep{peto1977epidemiology, leroi2003cancer, caulin2011peto, nunney2013real}. \citet{caulin2011peto} review numerous hypotheses to explain the observation, most of which involve cellular or tissue level cancer prevention or suppression. Our results indicate a similar effect at the level of a single population (humans), but show that the relationship is not flat as in the interspecific comparison, but rather an increasing, saturating function. Indeed, our analysis demonstrates that the saturation effect is due to different families of tissues, which characteristically differ in their ranges of total stem cell divisions. Most of the cancer types in this dataset occur at older ages and, as has been argued previously (e.g., \citet{nunney2013real}), such cancers would be shielded from present-day natural selection. Our hypothesis that natural selection for general cancer prevention and tissue-specific cancer prevention acting over millions of generations is consistent with the present day observations of occurrence shifted to older ages, yet maintaining the evolved protection mechanisms that reduce incidences at younger ages (during which the force of selection is expected to be greatest (e.g., \citet{hamilton1966moulding}).

Future research should extend Tomasetti and Vogelstein's dataset to other tissue types and cancer types within tissues (most notably high incidence cancers of the breast and prostate). Moreover, we need to identify possible tissue-specific mechanisms of cancer prevention to test the hypothesis that natural selection has influenced not only age related patterns in cancer incidence, but also tissue specific adaptations and cancer as a possible evolutionary constraint on tissue size.

\subsection*{Acknowledgements}

This work was supported by grants from the CNRS (EvoCan ANR-13-BSV7-0003-01) and INSERM (``Physique Cancer" (CanEvolve PC201306) to MEH. C\'{e}line Devaux, Vincent Devictor, Robert Gatenby, Pierre Gauz\`{e}re, Urszula Hibner, Pierre Martinez, Len Nunney and Christian Tomasetti provided helpful advice.

\bibliography{TVcommentary}

\section*{Appendix: methods and data}

All analyses were conducted in R \citep{Rlanguage} (except that Figure~\ref{fig:split} was generated in JMP \citep{JMPlanguage}). We obtained the dataset from the supplementary materials of \citet{Tomasetti2015} and assigned the following variable names: type (Cancer type in T\&V), risk (Lifetime cancer incidence), $s$ (Number of normal stem cells in tissue of origin), $d$ (Number of divisions of each stem cell per lifetime), lscd (Cumulative number of divisions of all stem cells per lifetime). We also added the factors ``morphology\_subset" and ``topography\_subset", as explained in our Results.

The multiple linear regression model (Equation~\ref{eqn:multi}) was implemented in R for all 31 cancer types as
\begin{verbatim}
> lm(log10(risk) ~ log10(d + 1) * log10(s), data = data)
\end{verbatim}
The models illustrated in Figures~\ref{fig:split}A and~\ref{fig:split}B were implemented as
\begin{verbatim}
> model1 <- lm(log10(risk) ~ log10(s), data = data)
> model2 <- lm(log10(risk) ~ log10(d+1), data = data)
> lm(resid(model1) ~ log10(data$d + 1))
> lm(resid(model2) ~ log10(data$s))
\end{verbatim}

The non-linear saturating model (Equation~\ref{eqn:logistic}, Figure~\ref{fig:logistic}A) was implemented for all 31 cancer types as
\begin{verbatim}
> logrisk <- log10(data$risk)
> loglscd <- log10(data$lscd)
> nls(logrisk ~ -log(a + exp(-loglscd - b)),
   + start = list(a = 0, b = -7),
   + lower = list(0, -20),
   + upper = list(100, -1),
   + algorithm = "port")
\end{verbatim}
This model was also run on the 24 cancer types not associated with a specific subpopulation (Figure~\ref{fig:logistic}B).

For the morphology subsets analysis (Equation~\ref{eqn:subsets}), when there was more than one risk value per cancer type (e.g. for lung cancer), we excluded the values that apply only to particular subpopulations. We also excluded subsets containing only one cancer type:
\begin{verbatim}
> data_morphology <- subset(data, (duplicated(morphology_subset)
   + | duplicated(morphology_subset, fromLast = TRUE))
   + & morphology_subset != "NA")
\end{verbatim}
We then ran the following regression model:
\begin{verbatim}
> lm(log10(risk) ~ log10(lscd) + morphology_subset,
   + data = data_morphology)
\end{verbatim}

Similarly for the topography subsets analysis:
\begin{verbatim}
> data_topography <- subset(data, (duplicated(topography_subset)
   + | duplicated(topography_subset, fromLast = TRUE))
   + & topography_subset != "NA")
> lm(log10(risk) ~ log10(lscd) + topography_subset,
   + data = data_topography)
\end{verbatim}

Finally, variation in cancer risk per stem cell division was modelled as
\begin{verbatim}
> data_topography$risk_per_div <-
   + log10(data_topography$risk) - log10(data_topography$lscd)
> lm(risk_per_div ~ topography_subset, data = data_topography)
\end{verbatim}

\begin{table}[p]
\begin{tabularx}{\textwidth}{X X}
\hline
Cancer type	&	Topographical subset	\\
\hline
Esophageal squamous cell carcinoma	&	C00-C15, C32 Mouth, Pharynx, Larynx, Esophagus	\\
Head and neck squamous cell carcinoma	&	C00-C15, C32 Mouth, Pharynx, Larynx, Esophagus	\\
Duodenum adenocarcinoma	&	C17-C20 Intestine, Rectum	\\
Small intestine adenocarcinoma	&	C17-C20 Intestine, Rectum	\\
Colorectal adenocarcinoma	&	C17-C20 Intestine, Rectum	\\
Hepatocellular carcinoma	&	C22 Liver	\\
Gallbladder non papillary adenocarcinoma	&	C23 Gallbladder	\\
Pancreatic endocrine (islet cell) carcinoma	&	C25 Pancreas	\\
Pancreatic ductal adenocarcinoma (acinar)	&	C25 Pancreas	\\
Lung adenocarcinoma (nonsmokers)	&	C34 Lung	\\
Osteosarcoma	&	C40-C41 Bone	\\
Osteosarcoma of the arms	&	C40-C41 Bone	\\
Osteosarcoma of the head	&	C40-C41 Bone	\\
Osteosarcoma of the legs	&	C40-C41 Bone	\\
Osteosarcoma of the pelvis	&	C40-C41 Bone	\\
Chronic lymphocytic leukemia	&	C42 Blood	\\
Acute myeloid leukemia	&	C42 Blood	\\
Basal cell carcinoma	&	C44 Skin	\\
Melanoma	&	C44 Skin	\\
Ovarian germ cell	&	C56 Ovary	\\
Testicular germ cell cancer	&	C62 Testis	\\
Glioblastoma	&	C71 Brain	\\
Medulloblastoma	&	C71 Brain	\\
Thyroid papillary/follicular carcinoma	&	C73 Thyroid	\\
Thyroid medullary carcinoma	&	C73 Thyroid	\\
\hline
\end{tabularx}
\caption{Topographical subsets assigned to cancer types in the T\&V dataset, according to the International Classification of Diseases for Oncology \citep{WHOinternational}.}
\label{tbl:subsets_top}
\end{table}

\begin{table}[p]
\begin{tabularx}{\textwidth}{X X}
\hline
Cancer type	&	Morphological subset	\\
\hline
Esophageal squamous cell carcinoma	&	M8070 Squamous cell carcinoma	\\
Head and neck squamous cell carcinoma	&	M8070 Squamous cell carcinoma	\\
Basal cell carcinoma	&	M8090 Basal cell carcinoma	\\
Gallbladder non papillary adenocarcinoma	&	M8140 Adenocarcinoma	\\
Duodenum adenocarcinoma	&	M8140 Adenocarcinoma	\\
Small intestine adenocarcinoma	&	M8140 Adenocarcinoma	\\
Colorectal adenocarcinoma	&	M8140 Adenocarcinoma	\\
Lung adenocarcinoma (nonsmokers)	&	M8140 Adenocarcinoma	\\
Thyroid papillary/follicular carcinoma	&	M8140 Adenocarcinoma	\\
Pancreatic endocrine (islet cell) carcinoma	&	M8150 Pancreatic endocrine	\\
Hepatocellular carcinoma	&	M8170 Hepatocellular carcinoma	\\
Thyroid medullary carcinoma	&	M8510 Medullary carcinoma	\\
Pancreatic ductal adenocarcinoma (acinar)	&	M8550 Acinar cell carcinoma	\\
Melanoma	&	M8720 Malignant melanoma	\\
Ovarian germ cell	&	M9060-9085 Germ cell tumours	\\
Testicular germ cell cancer	&	M9060-9085 Germ cell tumours	\\
Osteosarcoma	&	M9180 Osteosarcoma	\\
Osteosarcoma of the arms	&	M9180 Osteosarcoma	\\
Osteosarcoma of the head	&	M9180 Osteosarcoma	\\
Osteosarcoma of the legs	&	M9180 Osteosarcoma	\\
Osteosarcoma of the pelvis	&	M9180 Osteosarcoma	\\
Glioblastoma	&	M9440 Glioblastoma	\\
Medulloblastoma	&	M9470 Medulloblastoma	\\
Chronic lymphocytic leukemia	&	M9823 B-cell lymphocytic leukemia	\\
Acute myeloid leukemia	&	M9896 Acute myeloid leukaemia	\\
\hline
\end{tabularx}
\caption{Morphological subsets assigned to cancer types in the T\&V dataset, according to the International Classification of Diseases for Oncology \citep{WHOinternational}.}
\label{tbl:subsets_morph}
\end{table}

\end{document}